\documentclass[twocolumn]{aastex631}

\usepackage{color}

\shorttitle{AASTeX v6.3.1 Sample article}
\usepackage{float}
\usepackage{booktabs}
\usepackage{subfigure}
\usepackage{graphicx}
\usepackage{footnote}
\usepackage{multirow}
\usepackage{color}
\usepackage{amsmath}
\usepackage{cases}
\usepackage{appendix}
\graphicspath{{./}{figures/}}

\begin{document}

\title{Discovery of diffuse $\gamma$-ray emission in the vicinity of G213.0-0.6: Supernova remnant versus massive star-forming region} 

\author[0009-0003-4873-6770]{Yuan Li}
\affiliation{Tsung-Dao Lee Institute, Shanghai Jiao Tong University, Shanghai 201210, People's Republic of China}
\affiliation{School of Physics and Astronomy, Shanghai Jiao Tong University, Shanghai 200240, PRC}

\author[0000-0001-9745-5738]{Gwenael Giacinti}
\affiliation{Tsung-Dao Lee Institute, Shanghai Jiao Tong University, Shanghai 201210, People's Republic of China}
\affiliation{School of Physics and Astronomy, Shanghai Jiao Tong University, Shanghai 200240, PRC}
\affiliation{Key Laboratory for Particle Physics, Astrophysics and Cosmology (Ministry of Education) $\&$ Shanghai Key Laboratory for Particle Physics and Cosmology 800 Dongchuan Road, Shanghai, 200240, PRC }

\author[0000-0003-1039-9521]{Siming Liu}
\affiliation{School of Physical Science and Technology, Southwest Jiaotong University, Chengdu 610031, PRC}

\email{YuanLss17@sjtu.edu.cn}
\email{gwenael.giacinti@sjtu.edu.cn}
\email{liusm@swjtu.edu.cn}

\begin{abstract}
We report the discovery of high-energy $\gamma$-ray emission in the vicinity of G213.0-0.6, which is debated as a supernova remnant (SNR) or an ionized hydrogen (H$_{\rm{II}}$) region. Using 16-yr Pass 8 data from $\emph{Fermi}$ Large Area Telescope ($\emph{Fermi}$-LAT), we found three extended sources with different photon spectra in this region, which will label as SrcA, SrcB and SrcC. Among them, the $\gamma$-ray source SrcA with a log-parabola spectrum is spatially coincident with a star-forming region and several OB stars. The power-law spectra source SrcB is spatially coincident with a SNR radio shell. SrcC with a harder power-law photon spectrum is located outside of the radio shell structure. All of them are spatially coincident with a dense molecular cloud (MC) in the velocity range of 35 - 54 km s$^{-1}$. In this scenario, SrcB can be interpreted as the GeV counterpart of the SNR, and its $\gamma$-ray emission originates from the shock-cloud interaction. SrcA and SrcC originate from the escaped CRs illuminating nearby MC. For SrcA, another possibility is that the $\gamma$-ray emission originates from a young stellar cluster (YSC) associated with a star-forming region (SFR), however, the supporting evidence remains insufficient to draw a definitive conclusion. 
\end{abstract}


\section{Introduction} \label{sec:intro}

It is believed that Galactic cosmic rays (CRs) are accelerated by supernova remnants (SNRs) or young massive star clusters (YSC) in our Galaxy. In the former case, SNRs interacting with dense molecular clouds (MCs) are expected to be bright in the $\gamma$-ray band. Indeed, the $\gamma$-ray emissions from such systems have been detected by the Fermi Large Area Telescope (\emph{Fermi}-LAT), including W44 \citep{uchiyama2012fermi, peron2020gamma}, W28 \citep{aharonian2008discovery,hanabata2014detailed} and SNR G150.3+4.5 \citep{2024A&A...689A.257L}. The intense GeV $\gamma$-ray emissions from these SNRs are commonly considered to be from the decay of neutral pions generated in inelastic collisions between accelerated protons and the dense gas in Molecular cloud (MC). The derived $\gamma$-ray flux depends on the amount of nuclear CRs released and the diffusion coefficient in the interstellar medium \citep[ISM;][]{gabici2009,aharonian2004,marrero2008}. 

On the other hand, during the past ten years, an increasing number of observations indicate that the YSC associated with star-forming regions (SFR) is a significant class of Galactic cosmic rays factories \citep{2019NatAs...3..561A}. CRs could be efficiently accelerated by strong fast winds of young massive stars and shocks caused by the core-collapse SNe in clusters. Given that the SN blast wave may interact with fast star winds, it is anticipated that the acceleration efficiency and maximum particle energy will be improved in comparison to the typical values obtained for a single SNR shock \citep{2020SSRv..216...42B}. According to W30 \citep{2019ApJ...881...94L} and W40 \citep{2020A&A...639A..80S}, YSCs usually host dense molecular gas to drive their strong star formation. As a result, the hadronic interaction of accelerated CRs with the surrounding dense gas is a natural explanation for their $\gamma$-ray emission. Thus far, extended $\gamma$-ray structures from GeV to TeV bands have been detected in a number of such systems, including the Cygnus cocoon associated with the compact cluster Cygnus OB2 \citep{2011Sci...334.1103A}, Westerlund 1 \citep{2012A&A...537A.114A}, Westerlund 2 \citep{2018A&A...611A..77Y}, NGC 3603 \citep{2017A&A...600A.107Y}, and 30 Dor C \citep{2015Sci...347..406H}. Specifically, ultra high energy (UHE; $>$100TeV) $\gamma$-ray emission for the Cygnus cocoon was recently identified by LHAASO in the energy range up to a few PeV \citep{2024SciBu..69..449L}.

G213.0-0.6 (Hereafter G213) was suggested to be an extended radio faint shell-type SNR located in the anticentre region of Galactic plane \citep{2003A&A...408..961R,2019JApA...40...36G}, and no corresponding extended component was found in the $\gamma$-ray energy band before \citep{2016ApJS..224....8A}. No energetic pulsar was detected in this region, and its radio continuum spectral index ($S_\nu \propto \nu^{\alpha}$) measured by \citet{2024MNRAS.527.7355G} and \citet{2003A&A...408..961R} in this region is much softer than that of a typical PWN ($-0.3 \lesssim \alpha \lesssim 0$, \cite{2006ARA&A..44...17G}). These evidence suggest G213 is more compatible with a SNR scenario with $\alpha \lesssim -0.40$ \cite{2017yCat.7278....0G}. However, \citet{2024MNRAS.527.7355G} also argue another possibility that G213 might be just a H$_{\rm{II}}$ region. Recently gas work done by \citet{2017ApJ...836..211S} argue the revised distance of the SNR should be around 1 kpc, which was confirmed in \citet{2019MNRAS.488.3129Y} and \citet{2020ApJ...891..137Z} by optical and dust analysis. This means the prominent H$_{\rm{II}}$ region SH 2-284 \citep{1959ApJS....4..257S} has no physical relationship with SNR G213, because of its 4 kpc \citep{2011MNRAS.410..227C} or 4.5 kpc distance \citep{2015A&A...584A..77N}. However, combining with gas and $\gamma$-ray emission distribution, our results suggest that G213 might be a SNR with a similar distance to the ionized region. Furthermore, we found a $\gamma$-ray source whose origin might come from the escaped CR accelerated in SNR G213 and illuminating a nearby dense cloud, which is located around 4.4 kpc. At the same time, we also found another $\gamma$-ray source located within the projection region of ionized hydrogen SH 2-284, indicating the CRs might be accelerated by the massive stellar wind released by OB stars in associated star-forming region (SFR).

In this work, we conduct a comprehensive analysis of the GeV $\gamma$-ray emission in the vicinity of G213 region, utilizing 16 years of data from the \emph{Fermi}-LAT instrument. The photon events are selected in an energy range from 100 MeV to 1 TeV, and the results are presented in Sect. \ref{sec:2}. In Sect. \ref{sec:3}, we present the gas observation results of $^{12}$CO(J = 1-0) obtained from the Milky Way Imaging Scroll Painting (MWISP). Sect. \ref{sec:4} is dedicated to discussing potential origins of the $\gamma$-ray emission. Lastly, Sect. \ref{sec:5} provides our conclusions.

\section{\emph{Fermi}-LAT Data Reduction}\label{sec:2}

In the subsequent analysis, the standard LAT analysis software \emph{Fermitools} v2.2.0 is adopted, together with the \emph{Fermipy} v1.1.6 \citep{2017ICRC...35..824W} to quantitatively calculate the extension and position of extended sources. The latest Pass 8 data are collected from August 4, 2008 (Mission Elapsed Time 239557418) to August 4, 2024 (Mission Elapsed Time 744465605) to study the GeV emission around G213 region.
We select the "Source" event class together with instrumental response function ``P8R3$\_$SOURCE'' (evclass=128) and event type FRONT + BACK (evtype=3), with the standard data quality selection criteria $\tt (DATA\_QUAL > 0)  \&\& (LAT\_CONFIG == 1)$, and exclude the zenith angle over 90$\degr$ to avoid the earth limb contamination. The energies of events are cut between 1 GeV and 1 TeV are selected to do the spatial analysis. Additionally, events with energies between 100 MeV and 1 TeV are selected to do a more detailed spectral analysis. The analysis was performed in a  $14\degr\times14\degr$ region of interest (ROI) with the standard LAT analysis software ScienceTools. All sources listed in the incremental version of the fourth Fermi-LAT source catalog (4FGL-DR4;\citep{2020ApJS..247...33A,2023arXiv230712546B}) are adopted for the binned maximum likelihood analysis \citep{mattox1996likelihood}. The Galactic/isotropic diffuse background models (IEM, $\tt gll\_iem\_v07.fits$)/($\tt iso\_P8R3\_SOURCE\_V3\_v1.txt$ ) are adopted, all sources listed in the 4FGL-DR4 catalog are included in the background model, and all sources within $20\degr$ from the center of ROI and two diffuse backgrounds are included in the model, which is generated by the script make4FGLxml.py\footnote{\url{ http://fermi.gsfc.nasa.gov/ssc/data/analysis/user}}. The likelihood test statistic (TS) is adopted to calculate the significance of the $\gamma$-ray sources, which is defined as TS$= 2 (\ln\mathcal{L}_{1}-\ln\mathcal{L}_{0})$, where $\mathcal{L}_{1}$ and $\mathcal{L}_{0}$ are maximum likelihood values for the model with target source and without target source. Furthermore, the $\rm TS_{ext}$ is defined as $\rm {TS_{ext}} = 2(\ln\mathcal{L}_{\rm ext} - \ln\mathcal{L}_{\rm ps})$, where $\mathcal{L}_{\rm ext}$ and $\mathcal{L}_{\rm ps}$ represent the maximum likelihood values for the extended and point-like templates, respectively. This calculation considers only one additional free parameter introduced by the extended template, and the extension significance is approximately given by $\sqrt{\rm TS_{ext}}$ in units of $\sigma$.

\subsection{Morphological analysis}\label{sec:2.2}
\begin{figure*}
    \centering
    \includegraphics[trim={0 0.cm 0 0}, clip,width=0.49\textwidth]{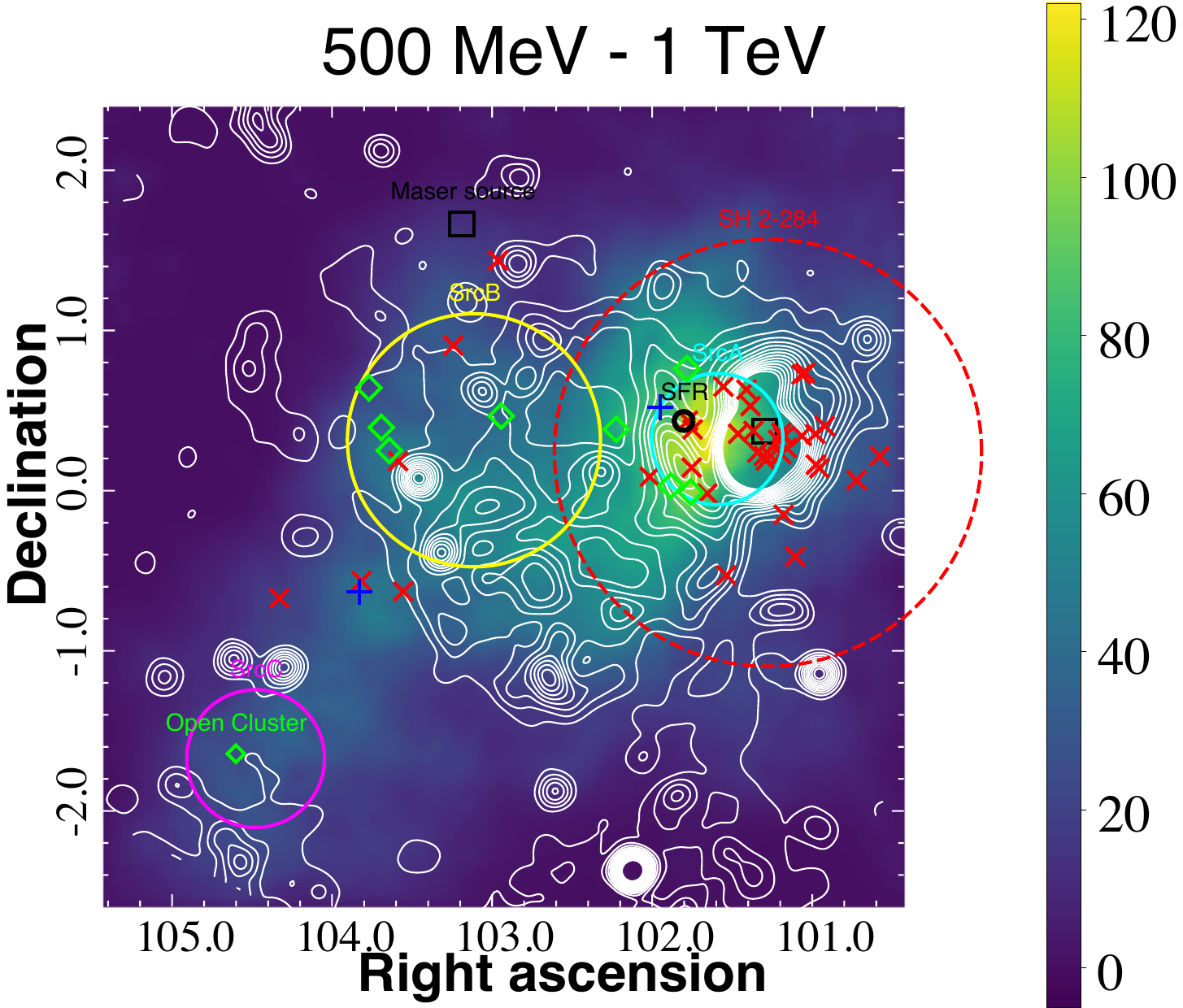} 
    \includegraphics[trim={0 0.cm 0 0}, clip,width=0.495\textwidth]{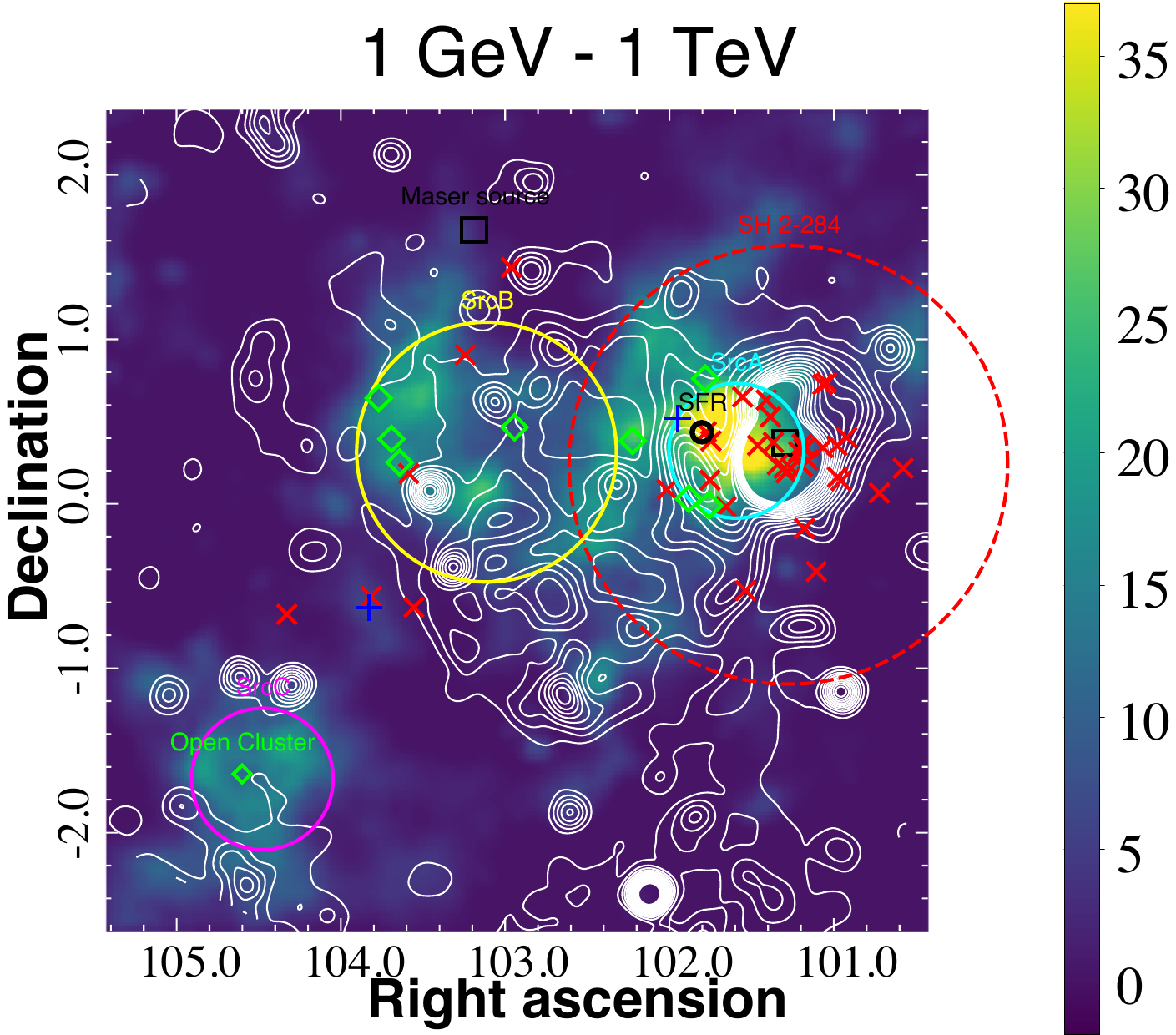} \\
    \includegraphics[trim={0 0.cm 0 0}, clip,width=0.499\textwidth]{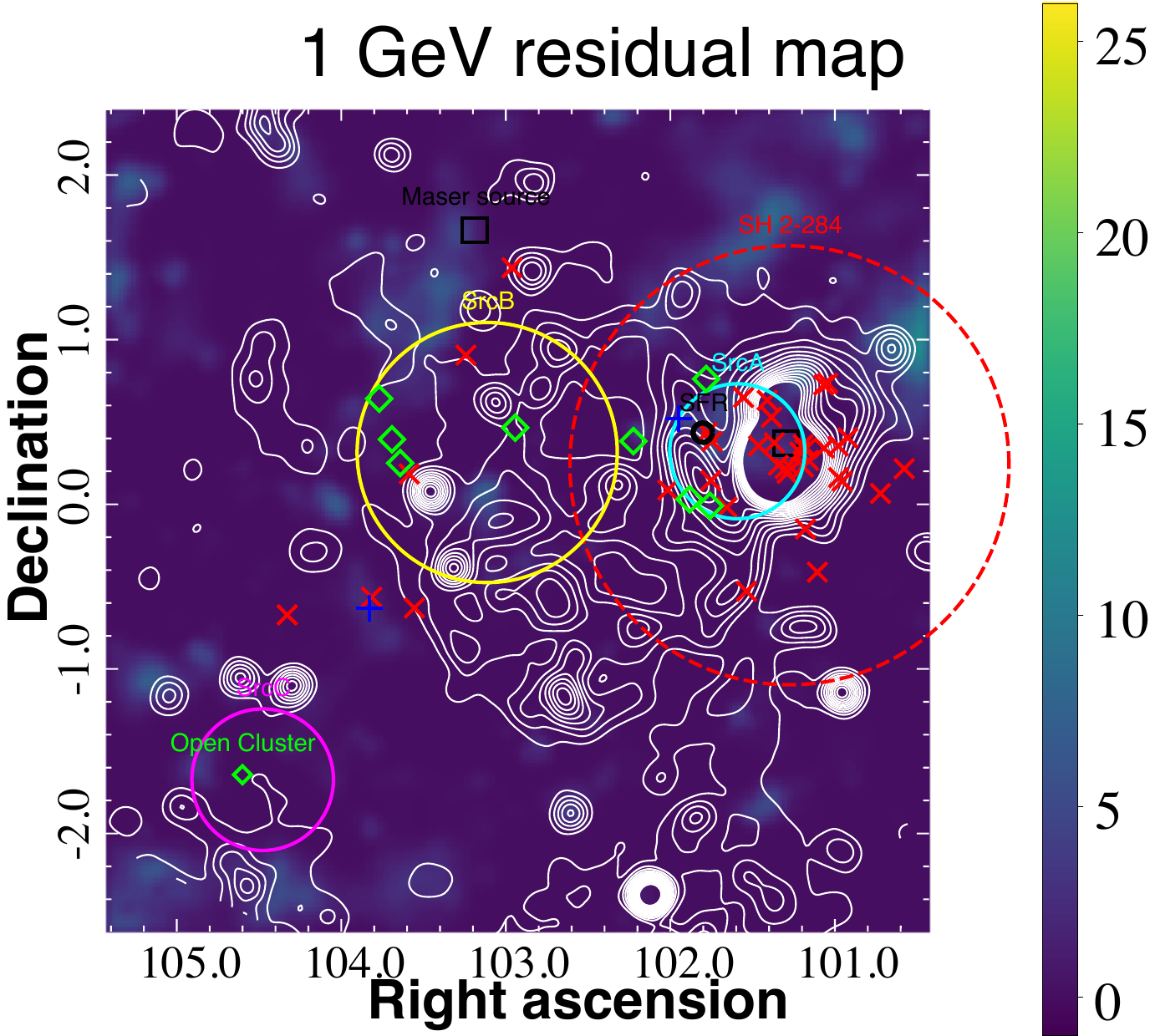}
    \includegraphics[trim={0 0.cm 0 0}, clip,width=0.495\textwidth]{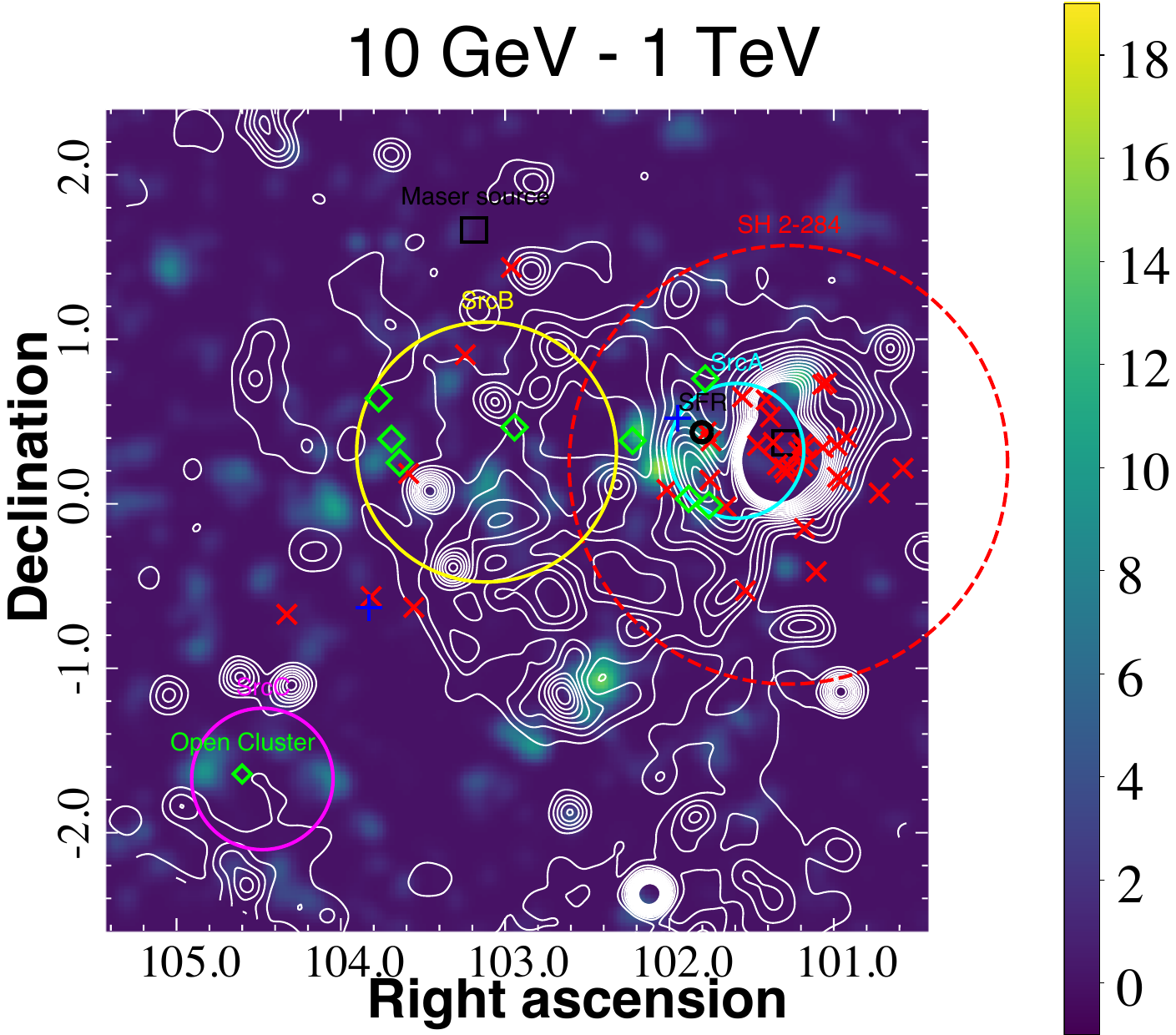}\\
    \caption{5$\degr$$\times$5$\degr$ TS maps in different energy bands. The two blue crosses show the point-like sources (4FGL J0647.7+0031 and 4FGL J0655.3-0037) in this region given by 4FGL-DR4. The cyan, yellow and magenta circles show the best-fit R$_{68}$ extension size for SrcA, SrcB and SrcC with uniform disk template, respectively. The white contours show the radio emission results from the Effelsberg 1.4-GHz survey data \citep{1997A&AS..126..413R}. The green diamonds and black squares show the positions of open clusters and maser sources from SIMBAD (\url{ http://simbad.u-strasbg.fr/simbad/sim-fbasic}) database. The black circle shows the position of star forming region [SUH2012] G212.064-00.739 \citep{2012A&A...542A...3S}. The red crosses represents the position of OB stars extracted from \citet{2017ApJ...836..211S}. The red dashed circle shows the extension of ionized hydrogen region SH 2-284 \citep{1976A&AS...25...25D}.}
   \label{fig:1}
 \end{figure*}

In the region around G213, two point-like $\gamma$-ray sources are listed in 4FGL-DR4 catalog (4FGL J0647.7+0031 and 4FGL J0655.3-0037), both of them do not have identified counterparts in others wavelength. We first created a 5$\degr \times$ 5$\degr$ region in the TS map with the command $gttsmap$ by subtracting the emission from the diffuse backgrounds and all 4FGL-DR4 sources (excepts 4FGL J0647.7+0031 and 4FGL J0655.3-0037) in the best-fit model using the events in the energy range of 500 MeV - 1TeV. Strong diffuse $\gamma$-ray emission appears shown as in the top left panel in Figure \ref{fig:1}. Considering a better point spread function (PSF) in the higher energy band as well as sufficient photon events, we only select events above 1 GeV to do further spatial analysis, and the generated TS map in this energy range is shown in the top right panel in Figure \ref{fig:1}.

In order to determine the best-fit spatial template above 1 GeV for the diffuse $\gamma$-ray emission in this region, we first added a single uniform disk template into the model (Model 2) to take place of the original two point-like source in 4FGL-DR4 model (Model 1), the derived best-fit position and 68\% error radius from \emph{Fermipy} is recorded as (R.A. = $102^{\circ}\!.483$, Dec. = $0^{\circ}\!.082$,  r$_{\rm 68}$ = $1^{\circ}\!.15$). Compared with the 4FGL-DR4 model (Model 1), the Model 2 is significantly improved by $\Delta$TS $\approx$ 74. After subtracting the $\gamma$-ray emission from Model 2, there is still a significant diffuse $\gamma$-ray emission located in the south-eastern direction, along with a strong $\gamma$-ray excess located in the western boundary of the disk template. Therefore, we additionally added a disk template into the model (Model 3) and repeat the fitting process. The derived best-fit position and 68\% error radius are recorded as (R.A. = $101^{\circ}\!.594$, Dec. = $0^{\circ}\!.322$,  r$_{\rm 68}$ = $0^{\circ}\!.41$) and (R.A. = $103^{\circ}\!.113$, Dec. = $0^{\circ}\!.314$,  r$_{\rm 68}$ = $0^{\circ}\!.79$), respectively. After subtracting the contribution from Model 3, the south-eastern $\gamma$-ray excess still exists, thus we further added a point-like source into the model (Model 4) and search for its spatial parameters with command $gtfindsrc$. The best-fit results is recorded as (R.A. = $104^{\circ}\!.532$, Dec. = $-1^{\circ}\!.813$,  r$_{\rm 68}$ = $0^{\circ}\!.08$). Then we calculated the $\rm TS_{ext}$ value of this point-like source to test the hypothesis that it is extended (Model 5). The $\rm TS_{ext}$ is calculated as 43, which means that the point-like source hypothesis is rejected around $6.5\sigma$. The best-fit position and 68\% error radius for this south-eastern extended source are (R.A. = $104^{\circ}\!.477$, Dec. = $-1^{\circ}\!.673$,  r$_{\rm 68}$ = $0^{\circ}\!.43$). Then we attempted to replace the disk with a 2D-Gaussian template, and the derived results are recorded in Model 6 to Model 9 in Table \ref{tab:1}. In addition to the likelihood value, we also used the Akaike information criterion (AIC) value \citep{1974AIC} to evaluate the quality of the template, where AIC is defined as AIC$ = 2k - 2\ln\mathcal{L}$ , where $k$ is the number of degrees of freedom of the model and $\mathcal{L}$ is the likelihood value. The model with the minimum AIC value is preferred, thus Model 5 is selected for the further spectral analysis. The measured results are summarized in Table \ref{tab:1}.

\begin{table*}
\centering
\caption{\textbf{Spatial templates tested for the GeV $\gamma$-ray emission}}
\begin{tabular}{ccccccc}
\hline \hline
Morphology($>$1GeV)  &R.A., Decl &Best$-$fit Extension (R$_{\rm 68}$)& -log(Likelihood)&Ndf$^{\,\,\text{a}}$&$\Delta${AIC}$^{\,\,\text{b}}$\\
\midrule
\multirow{2}*{Model 1 (4FGL-DR4)}&$101^{\circ}\!.949$, $0^{\circ}\!.521$&\multirow{2}*{$-$}&\multirow{2}*{46436}&\multirow{2}*{8}&\multirow{2}*{0} \\
~&$103^{\circ}\!.828$, $-0^{\circ}\!.631$&~&~&~&~&~\\
\hline
Model 2 (Single Disk)&$102^{\circ}\!.483, 0^{\circ}\!.082$&$1^{\circ}\!.15 \pm0^{\circ}\!.15$&46399&5&-47\\
\hline
\multirow{2}*{Model 3 (Two Disks)}&SrcA: $101^{\circ}\!.594, 0^{\circ}\!.322$&$0^{\circ}\!.41 \pm0^{\circ}\!.06$&\multirow{2}*{46378}&\multirow{2}*{10}&\multirow{2}*{-54}\\
~&SrcB: 103.113$\degr$, 0.314$\degr$&$0.79^{\circ}\! \pm0^{\circ}\!.10$&~&~&~\\
\hline
\multirow{3}*{Model 4 (Two Disks + point)}&\multirow{3}*{$-$}&\multirow{3}*{$-$}&\multirow{3}*{46362}&\multirow{3}*{14}&\multirow{3}*{-62}\\
~&~&~&~&~&~&~\\
~&SrcC: $104^{\circ}\!.532, -1^{\circ}\!.813$&~&~&~&~\\
\hline
\multirow{3}*{Model 5 (Three Disks)}&~&$0^{\circ}\!.41 \pm0^{\circ}\!.05$&\multirow{3}*{46341}&\multirow{3}*{15}&\multirow{3}*{-81}\\
~&$-$&$0^{\circ}\!.79 \pm0^{\circ}\!.06$&~&~&~\\
~& SrcC: $104^{\circ}\!.477, -1^{\circ}\!.673$&$0^{\circ}\!.43 \pm0^{\circ}\!.07$&~&~&~\\
\hline
Model 6 (2D Gaussian)&$102^{\circ}\!.561$, $0^{\circ}\!.112$&$1^{\circ}\!.13 \pm0^{\circ}\!.14$&46404&5&-38\\
\hline
\multirow{2}*{Model 7 (Two Gaussians)}&SrcA: $101^{\circ}\!.701$, $0^{\circ}\!.449$&$0^{\circ}\!.39 \pm0^{\circ}\!.05$&\multirow{2}*{46380}&\multirow{2}*{10}&\multirow{2}*{-52}\\
~ &SrcB: $103^{\circ}\!.526$, $0^{\circ}\!.365$&$0^{\circ}\!.72 \pm 0^{\circ}\!.08$&~&~&~\\
\hline
\multirow{3}*{Model 8 (Two Gaussians + point)}&\multirow{3}*{$-$}&\multirow{3}*{$-$}&\multirow{3}*{46395}&\multirow{3}*{14}&\multirow{3}*{-57}\\
~&~&~&~&~&~\\
~&SrcC: $104^{\circ}\!.344$, $-1^{\circ}\!.522$&~&~&~&~\\
\hline
\multirow{3}*{Model 9 (Three Gaussians)}&\multirow{3}*{$-$}&$0^{\circ}\!.38 \pm0^{\circ}\!.06$&\multirow{3}*{46343}&\multirow{3}*{15}&\multirow{3}*{-79}\\
~&~&$0^{\circ}\!.92 \pm0^{\circ}\!.05$&~&~&~\\
~&SrcC: $104^{\circ}\!.535$, $-1^{\circ}\!.696$&$0^{\circ}\!.39 \pm0^{\circ}\!.05$&~&~&~\\
\hline
\end{tabular}
\label{tab:1}
\\
{{\bf Notes.} $^{(a)}$ Number of degrees of freedom. $^{(b)}$ Calculated with respect to Model 1.}
\end{table*}

\subsection{Spectral analysis}\label{sec:2.2}

Now that the best-fit three disk template (Model 5) including SrcA/B/C in this region is determined, we adopt simple power-law (PL; dN/dE $\propto$ E$^{-\alpha}$) and logparabola (LogPb; dN/dE $\propto$ E$^{\rm -(\alpha+\beta log(E/E_{\rm b}))}$) functions to search for the best-fit spectra for each source. The measured results are summarized in Table \ref{tab:2}. For SrcA, compared with the single PL assumption, the LogPb spectrum shows a notable improvement of the TS value. This improvement can be quantified as $\rm{TS_{curve}}$, defined as $\rm{TS_{curve}}$=$2(\ln\mathcal{L}_{\rm BPL}-\ln\mathcal{L}_{\rm PL})$\citep{abdollahi2020a}. The obtained value of 20.7 corresponds to a significance level of $\sim$ 4.6 $\sigma$ with only one additional free parameter. Thus we conclude that there is an energy spectral break at $\sim$230 MeV in the SrcA spectrum. For SrcB and SrcC, compared with the single PL assumption, the TS value does not change significantly with LogPb spectrum, and the significance improvement is smaller than 9.0 (corresponding to $\sim$ 3 $\sigma$), indicating the single PL is enough to describe these two spectrum. Furthermore, to derive the spectral energy distributions (SEDs) of both sources, we divided the events in the 100 MeV - 1 TeV energy range into twelve logarithmically equal intervals and repeated the same likelihood fitting analysis for each interval. The normalizations of all sources are left free, and the spectral indices and energy break are fixed to their best-fit value. For bins with TS values less than 5.0, we give upper limits calculated with 95\% confidence level using a Bayesian method \citep{helene1983}.

\begin{table*}
    \caption{Parameters of the best-fit spectral models in the energy range of 100 MeV - 1 TeV} 
    \centering
    \begin{tabular}{lccccccc}
    \hline
Source& Spectral type&\textbf{$\Gamma$}1($\alpha$)&\textbf{$\Gamma$}2($\beta$)&E$_b$(GeV)&Photon flux(photon cm$^{-2}$ s$^{-1}$)&Degrees of Freedom&$\rm{TS_{curve}}$ \\
\hline
\multirow{2}*{SrcA}&PL&2.54$\pm$0.16&$-$&$-$&(2.69$\pm$0.33)$\times$10$^{-8}$&2&0.0 \\
~&LogPb& 1.82$\pm$0.17 & 0.18$\pm$0.06 & 0.23$\pm$0.15 & (2.43$\pm$0.39)$\times$10$^{-8}$ &3& 20.7 \\
\hline
\multirow{2}*{SrcB}&PL&2.30$\pm$0.12&$-$&$-$&(2.31$\pm$0.38)$\times$10$^{-8}$&2&0.0 \\
~&LogPb& 2.12$\pm$0.34 & 0.09$\pm$0.05 & 0.84$\pm$0.21 & (2.23$\pm$0.8)$\times$10$^{-8}$ &3& 3.4\\
\hline
\multirow{2}*{SrcC}&PL&2.17$\pm$0.19&$-$&$-$&(0.93$\pm$0.29)$\times$10$^{-8}$&2&0.0 \\
~&LogPb& 1.85$\pm$0.18 & 0.16$\pm$0.06 & 1.06$\pm$0.42 & (0.54$\pm$0.07)$\times$10$^{-8}$ &3& 6.5 \\
\hline
    \label{tab:2}
    \end{tabular}
\end{table*}

\begin{figure*}
    \centering
    \includegraphics[trim={0 0.cm 0 0}, clip,width=0.32\textwidth]{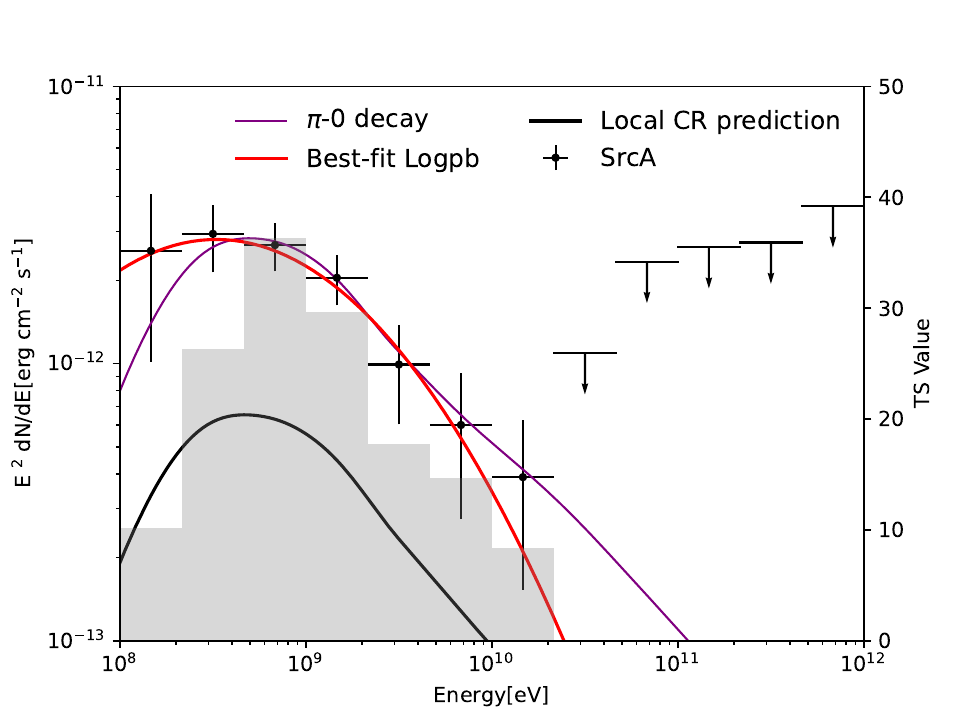}
    \includegraphics[trim={0 0.cm 0 0}, clip,width=0.32\textwidth]{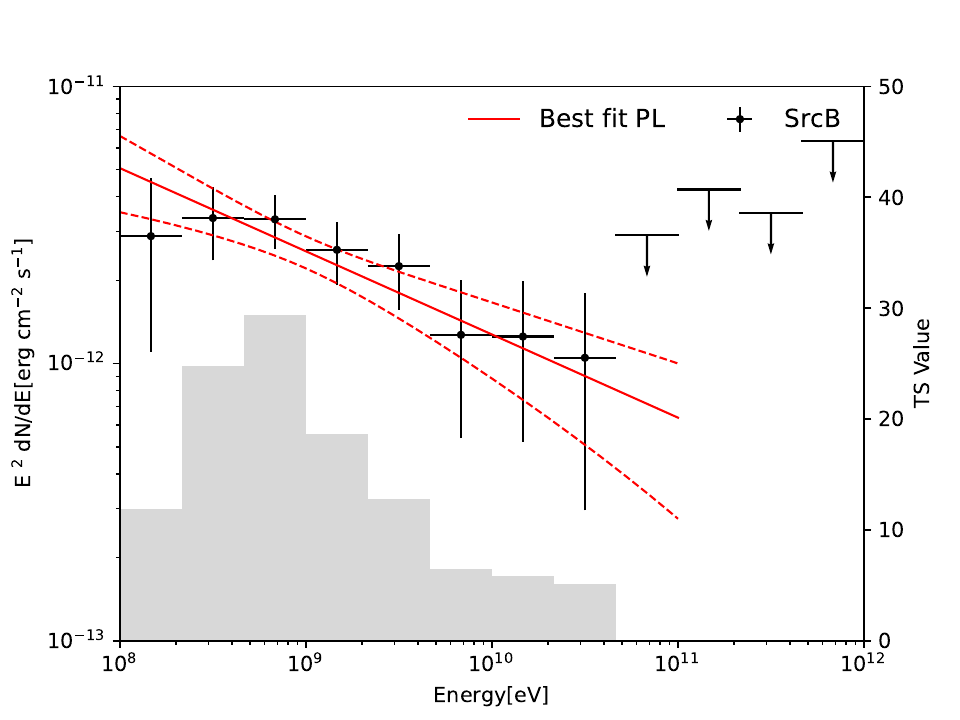}
    \includegraphics[trim={0 0.cm 0 0}, clip,width=0.32\textwidth]{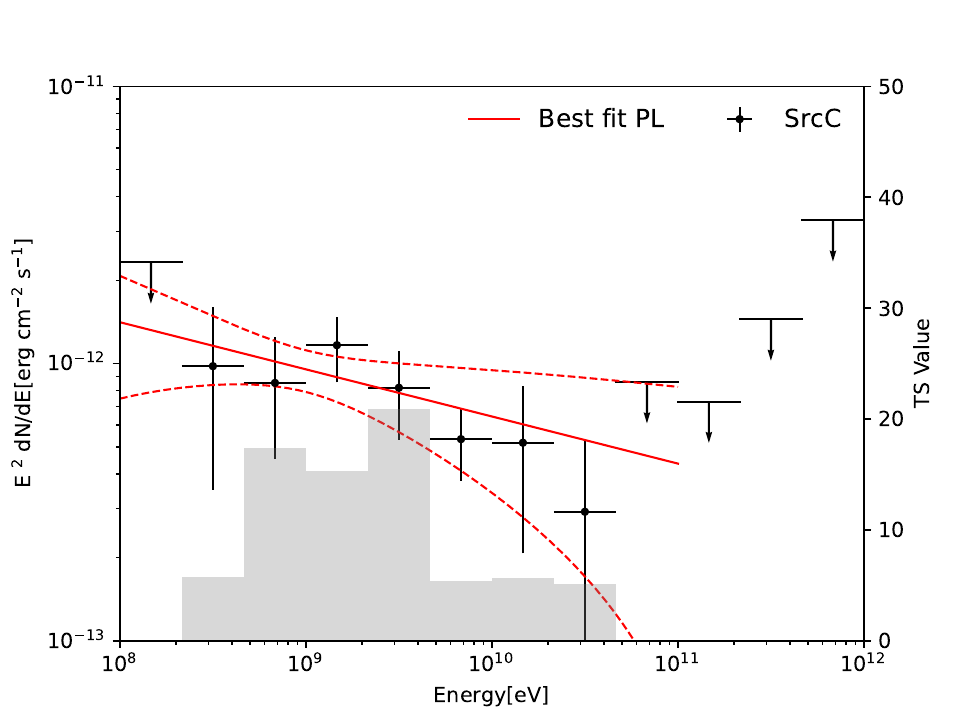}\\
    \caption{The SEDs of SrcA (left), SrcB (middle) and SrcC (right). The black data points are derived by Fermi-LAT in the energy range of 100 MeV - 1 TeV. The black arrows indicate the 95$\%$ upper limits and the grey histograms show the TS value for each energy bin. The red curve indicates the best-fit LogPb spectrum for SrcA. The red solid and dashed lines show the best-fit PL with 1 $\sigma$ statistical errors for SrcB and SrcC. The solid black line in the left panel represents the predicted $\gamma$-ray emission assuming that the CR spectra therein is the same as measured locally by AMS-02 \citep{2015PhRvL.115u1101A}. The magenta solid line represents a hadronic model assuming a power-law proton spectrum for SrcA (see Sect. \ref{sec:4.2}).}
   \label{fig:spectra}
 \end{figure*}

\section{Gas observations}\label{sec:3}
To trace the H$_{\rm 2}$, we make use of the data from the Milky Way Imaging Scroll Painting (MWISP \footnote{\url{http://english.dlh.pmo.cas.cn/ic/}}) project with high resolution CO survey along the Northern Galactic plane with the Purple Mountain Observatory (PMO) 13.7 m telescope \citep{2019ApJS..240....9S}. To determine the column density of H$_{\rm 2}$ in this region, we employ a conversion factor $X_\mathrm{CO}=2\times10^{20} \ \rm{cm^{-2} \ K^{-1} \ km^{-1} \ s}$ \citep{bolatto2013,2001ApJ...547..792D}. Using this factor, the column density $N_\mathrm{H_2}$ is calculated as $N_\mathrm{H_2} = X_\mathrm{CO} \times W_\mathrm{CO}$. Consequently, the mass of the molecular complex can be derived from the $W_\mathrm{CO}$:
\begin{equation}\label{eq:massco}
 M={\mu m_\mathrm{H}} D^2 \Delta\Omega_\mathrm{px} X_\mathrm{CO} {\sum_\mathrm{px}} W_\mathrm{CO} \propto N_\mathrm{H_2},
\end{equation}
In this formula, $\mu$ is set to 2.8, reflecting a relative helium abundance of 25$\%$. The mass of a hydrogen nucleon is denoted as $m_\mathrm{H}$. The solid angle subtended by each pixel is given by $\Delta\Omega_\mathrm{px}$. The term ${\sum_\mathrm{px}} W_\mathrm{CO}$ accounts for the velocity binning of the data cube. It is calculated by summing the map content for the pixels within the target sky region and the specified velocity range, and then scaling by the velocity bin size.

\begin{figure*}
     \centering
    \includegraphics[trim={0 0.cm 0 0}, clip,width=0.49\textwidth]{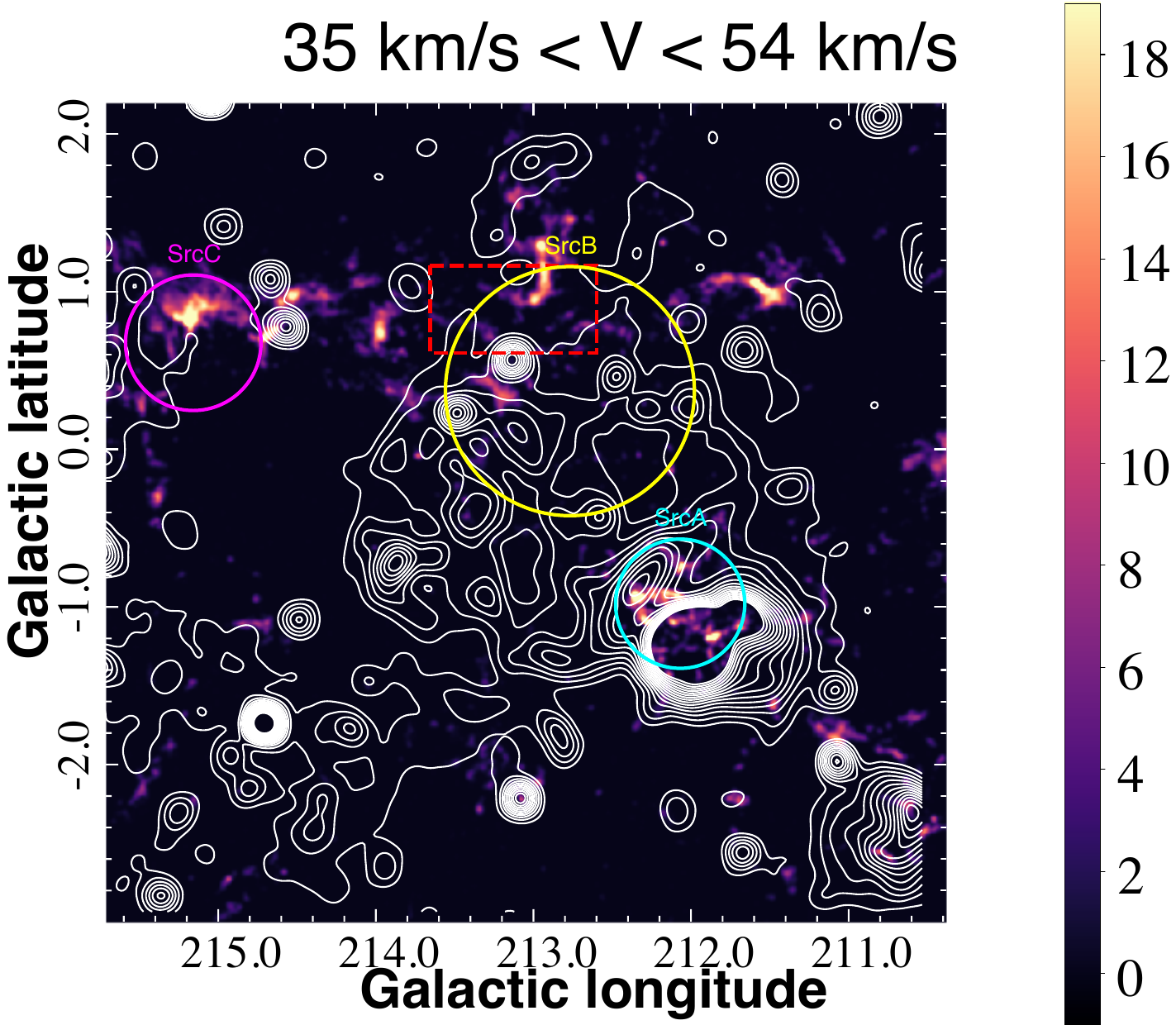}
    \includegraphics[trim={0 0.cm 0 0}, clip,width=0.49\textwidth]{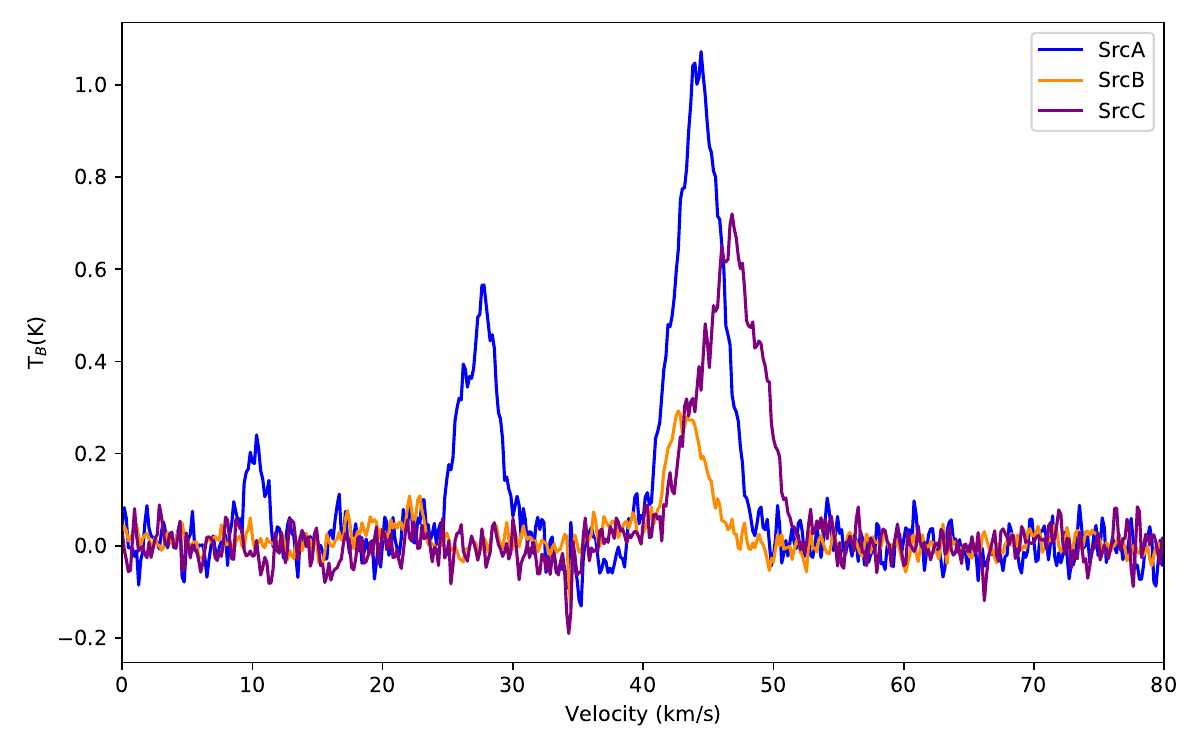}
    \caption{Left: Integrated $^{12}$CO(J = 1-0) emission intensity (K km s$^{-1}$) toward SNR G213 in the velocity range of 35 - 54 km s$^{-1}$. The white contours, the cyan, yellow and magenta circles are the same as in Figure \ref{fig:1}. The red box marks the region of highest gas density within SrcB. Right: $^{12}$CO(J = 1-0) spectra of gas inside each source region.}
\label{fig:3}
\end{figure*}

As shown in Figure \ref{fig:3}, SrcA and SrcC coincide well with dense regions of the gas distribution, while for SrcB, the spatial coincidence between the $\gamma$-ray emission and the dense gas is primarily observed in the northern part of the SNR radio shell. For the gas in the interval of 35 - 54 km s$^{-1}$, corresponding to the MCs at V$_{\rm LSR}$ = 42 - 48 km s$^{-1}$ \citep{2017ApJ...836..211S}, leads to a $\sim$ 4.4 kpc distance if $\Sigma-D$ relationship adopted \citep{2014ApJ...783..130R}. Since the radius for SrcA is measured as $\theta$ = 0.41 $\degr$, the total gas mass within SrcA region is estimated to be about $\rm{M_A} = 4.2\times 10^{4}d_{4.4}^{2} \ M_{\odot}$. Assuming a spherical geometry of the gas distribution, we estimate the volume to be $\rm{V_A} = {{4\pi \over3}R^3}$, here R = d $\times$ $\theta$, the average $\rm H_2$ cubic density in this region is about $\rm{n_{A}}$ = $\rm 13.1 d_{4.4}^{-1} \ cm^{-3}$. Similarly, the total gas mass within SrcB region ($\theta$ = 0.79 $\degr$) is estimated to be about $\rm{M_B} = 6.1\times 10^{4}d_{4.4}^{2} \ M_{\odot}$, corresponding to $\rm{n_{B}}$ = $\rm 3.5d_{4.4}^{-1} \ cm^{-3}$. For SrcC region ($\theta$ = 0.43 $\degr$), the total gas mass is estimated to be about $\rm{M_C} = 3.3\times 10^{4}d_{4.4}^{2} \ M_{\odot}$, corresponding to $\rm{n_C}$ = $\rm 8.4d_{4.4}^{-1} \ cm^{-3}$.  We also note that the highest gas density within SrcB is found in its northern part, where there is a spatial coincidence with several OB stars (shown as the red crosses in Figure \ref{fig:1}). In this northern region marked by the red box in Figure \ref{fig:3}, the gas density reaches approximately $\sim \rm 9d_{4.4}^{-1} \, cm^{-3}$. 

\section{Discussion of the possible origins of the $\gamma$-ray emission}\label{sec:4}
\subsection{SNR-MC scenario at the distance of 4.4 kpc}
\begin{figure*}
    \centering
    \includegraphics[trim={0 0.cm 0 0}, clip, width=0.32\textwidth]{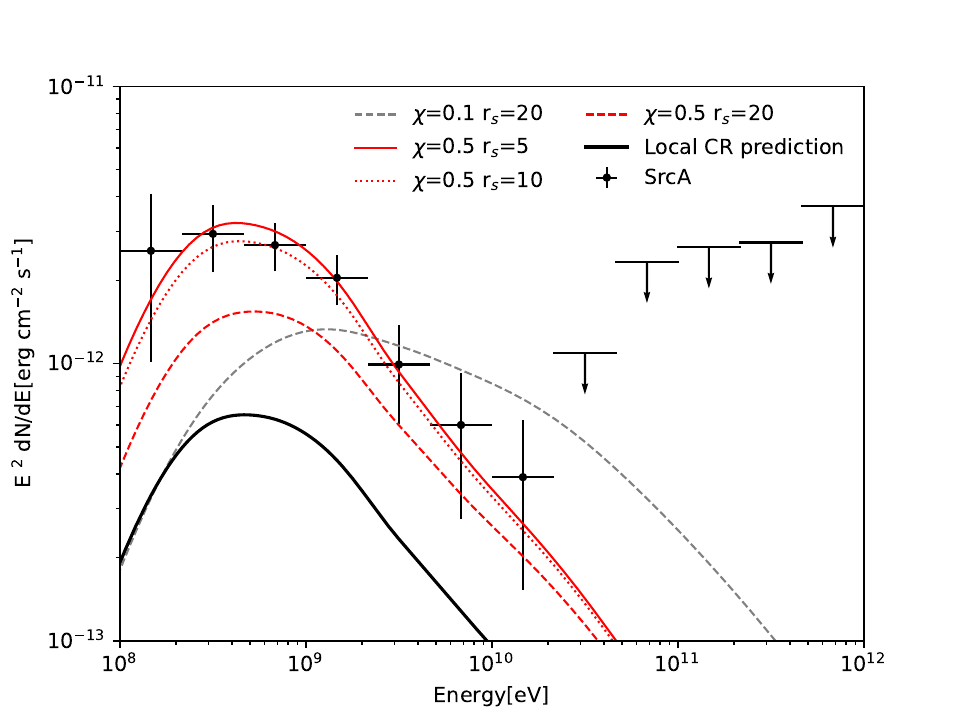}
    \includegraphics[trim={0 0.cm 0 0}, clip, width=0.32\textwidth]{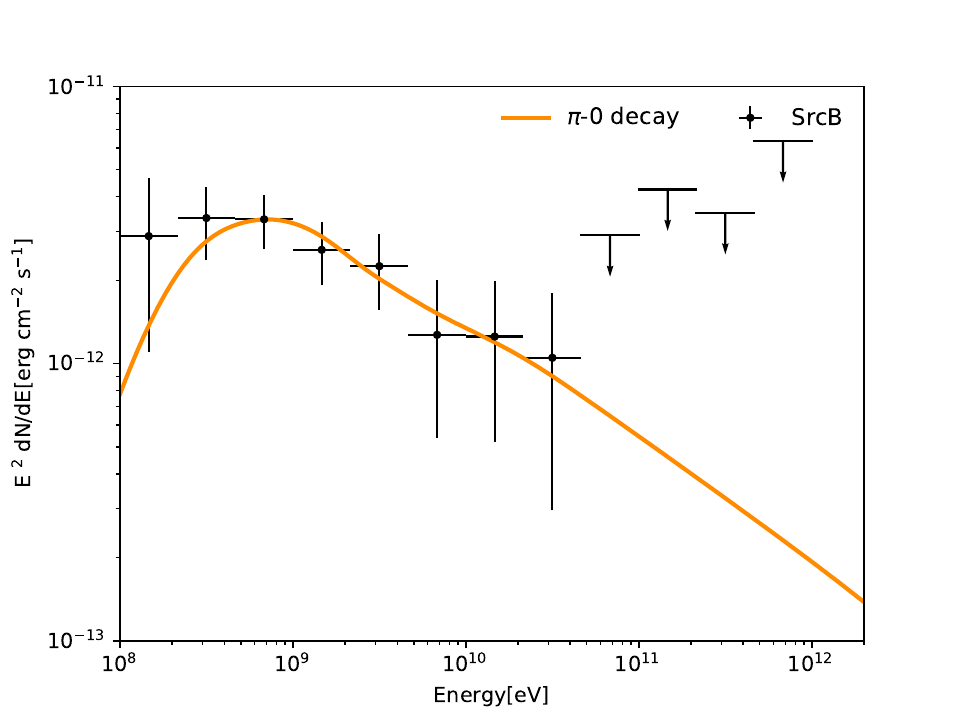}
    \includegraphics[trim={0 0.cm 0 0}, clip, width=0.32\textwidth]{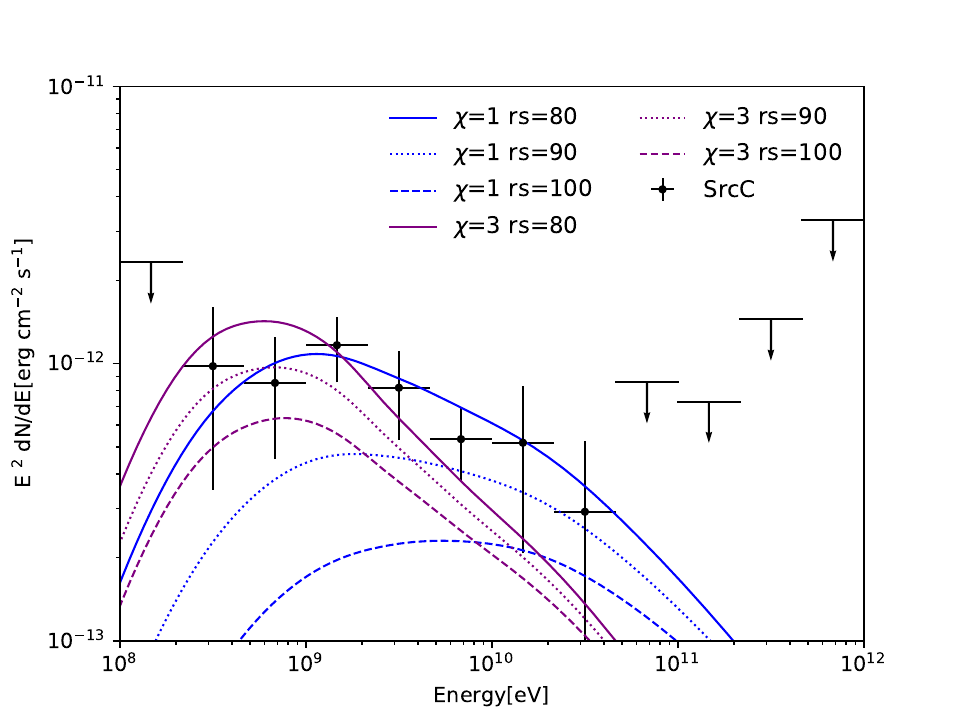}
    \caption{Hadronic model of the $\gamma$-ray spectra for SrcA (left), SrcB (middle) and SrcC (right). The black solid line in the left panel is the same as in Figure \ref{fig:spectra}. The solid, dotted and dashed fit lines for SrcA and SrcC show different parameters scenarios. The orange solid line in middle panel shown the hadronic model for a proton spectrum with an index of 2.5 and a cutoff energy of 500 TeV.}
    \label{fig:4}
\end{figure*}

In principle, it is possible that some $\gamma$-ray excesses arise because of the imperfect modeling of the Galactic diffuse $\gamma$-ray background. In particular, we note that the H$_{\rm II}$ component of the gas is not taken into account in the $Fermi$ diffuse background models \citep{2016ApJS..223...26A}. However, as the black solid line shows in Figure \ref{fig:spectra} and Figure \ref{fig:4}, the predicted $\gamma$-ray flux for the H$_{\rm II}$ gas (assuming that the CR spectra therein are the same as the local measurement \citep{2015PhRvL.115u1101A}) is significantly lower than the observed $\gamma$-ray flux. Thus, the derived $\gamma$-ray flux from this region (SrcA) cannot be caused by uncertainties in the modeling of the diffuse background. 

In Section \ref{sec:3}, we showed a good spatial coincidence between the molecular cloud and the two GeV sources, SrcA and SrcC, suggesting that the $\gamma$-ray emission may originate from hadronic $\pi^0$ decay. In this scenario, CRs that have escaped from the SNR shock surface illuminate the nearby MC gas, producing the observed $\gamma$-ray emission. In this case, the injected proton spectrum for SrcA and SrcC should be same as the protons spectrum of SNR G213, and the derived escape spectrum in Zone A and Zone C (corresponding to the MC located in SrcA and SrcC regions) might be influenced by the model parameters, such as the escape distance between the SNR and the target gas, the diffusion coefficient, etc. Here we adopted the method outlined by \citet{2020ApJ...897L..34L} to constrain the parameters space in the model.   

For simplicity, we consider a scenario where protons are injected instantaneously into two uniformly emitting regions Zone A and Zone C, approximately 11000 years ago \citep{2017ApJ...836..211S} and the spectrum of the injected protons can be characterized as a power-law with exponential cutoff:
\begin{eqnarray}
Q(E) = {Q_0} E^{-\Gamma} \exp \left(- \frac{E}{E_{\rm p, cut}} \right).
\label{eq:p_spectra}
\end{eqnarray}
\\
Here $\Gamma$ and $E_{\rm p, cut}$ are the spectral index and the cutoff energy of protons, respectively. Considering the $\gamma$-ray spectrum of Src B and its spatial coincidence with SNR G213, the injected spectrum is suggested to be $\Gamma$ = 2.5 to match with GeV spectrum, and the cutoff energy of protons is assumed to be $E_{\rm p, cut}$ = 500 TeV. The corresponding hadronic model for this proton spectrum is shown in the middle panel of Figure \ref{fig:4}.

The distribution of the escaped protons in the emission region can be calculated following the method outlined by \citet{2012MNRAS.419..624T,2020ApJ...897L..34L,2023ApJ...945...21L}:
\begin{equation}
    N_p(E,t)=\frac{Q(E)}{[4 \pi D(E) T]^\frac{3}{2}}  \exp\left(\frac{-r_{\rm s}^2}{4 D(E) T}\right) \label{equation:3}
\end{equation}
The diffusion coefficient is assumed to be uniform, following the form $D(E) = \chi D_0 (E/E_0)^\delta$ for $E > E_0$, where $D_0 = 1 \times 10^{28}$ cm$^2$ s$^{-1}$ at $E_0 = 10$ GeV and $\delta = 1/3$, consistent with Kolmogorov turbulence \citep{2006ApJ...642..902P,2013A&ARv..21...70B}. Due to projection effects, the actual distance between the gas complex and the SNR remains uncertain \citep{2023ApJ...953..100L}, here we adopted r$_{\rm s}$ as a free parameter representing the distance between the injection site and the illuminated molecular clouds. With the injected source spectrum defined by $Q(E) \propto E^{-\Gamma}$ and $D(E) \propto E^\delta$, equation \ref{equation:3} reveals that $N_p(E)$ will show a lower energy spectral cutoff at $E_{p, \rm break}$ when $\sqrt{4D(E_b)T} \simeq r_{\rm s}$. Correspondingly, $N_p(E)$ will follow $N_p(E) \propto E^{-\left(\Gamma+\frac{3}{2}\delta\right)}$. The total energy of injected protons is denoted as W$_{\rm inj}$ = $\eta$ E$_{\rm SN}$, where $\eta$ is the efficiency of converting kinetic energy into accelerated protons, typically valued at 0.1. The kinetic energy of the SNR, E$_{\rm SN}$ is generally taken to be 10$^{51}$erg \citep{2013A&ARv..21...70B}. Additionally, the correction factor $\chi$ for the diffusion coefficient is also set as a free parameter to check if the diffusion in SrcA and SrcC directions is isotropic. The corresponding $\gamma$-ray fluxes produced in the emission zone are then calculated with the $\emph{naima}$ package \citep{zabalza2015naima}.

In Figure \ref{fig:4}, for SrcA, the resulting hadronic $\gamma$-ray flux with $\chi$=0.5 and r$_{\rm s}$ = 5 pc or 10 pc can well explain the observational data, which is shown as the solid red line in the left panel, where the proton energy above 1 GeV is calculated as W$_{\rm p}$ = 2.15$\times$ 10$^{49}$ (M$_{\rm {A}}$/4.2$\times 10^{4}$ ${\rm M_{\odot}}$)$^{-1}$ erg. Furthermore, we noticed that r$_{\rm s}$ = 10 pc (red dotted line) can also match with the GeV data points, indicating the real distance between Zone A and the SNR should be very close. The proton energy above 1 GeV is calculated there as W$_{\rm p}$ = 1.79$\times$ 10$^{49}$ (M$_{\rm {A}}$/4.2$\times 10^{4}$ ${\rm M_{\odot}}$)$^{-1}$ erg. Additionally, we notice that as r$_{\rm s}$ increases, the derived $\gamma$-ray flux decreases dramatically, and the results for the r$_{\rm s}$ = 20 pc case, shown as the red dashed line, can be ruled out. We also tested the case with a small $\chi$ value for SrcA, which would lead to a much harder spectrum in the GeV band, as shown by the gray dashed line, and the total energy of protons above 1 GeV in this case is calculated as W$_{\rm p}$ = 5.36$\times$ 10$^{48}$ (M$_{\rm {A}}$/4.2$\times 10^{4}$ ${\rm M_{\odot}}$)$^{-1}$ erg. For SrcB, to explain its $\gamma$-ray emission, we adopted here the proton distribution following the formula $Q(E)$, and a rough ambient gas mass value in this region of $\rm{M_B}$ = $6.1\times 10^{4}d_{4.4}^{2} \ M_{\odot}$, while the total energy of protons above 1 GeV is estimated to be about W$_{\rm p,B}$ = 5.9$\times$ 10$^{49}$ (M$_{\rm {B}}$/6.1$\times 10^{4}$ ${\rm M_{\odot}}$)$^{-1}$ erg. For SrcC, since the GeV data points are located between the blue solid line and blue dashed line, the parameters r$_{\rm s}$ should be constrained between 80 pc and 90 pc. The energies of the protons above 1 GeV for r$_{\rm s}$ = 80 pc and 90 pc are calculated as 6.80$\times$ 10$^{48}$ (M$_{\rm {C}}$/3.3$\times 10^{4}$ ${\rm M_{\odot}}$)$^{-1}$ erg and 2.84$\times$ 10$^{48}$ (M$_{\rm {C}}$/3.3$\times 10^{4}$ ${\rm M_{\odot}}$)$^{-1}$ erg, respectively. We also noticed that the magenta fit lines lie slightly below the measured GeV data points, suggesting a constraint on the diffusion coefficient in Zone C (For the solid magenta line, the proton energy above 1 GeV is calculated as 1.17$\times$ 10$^{49}$ (M$_{\rm {C}}$/3.3$\times 10^{4}$ ${\rm M_{\odot}}$)$^{-1}$ erg). In particular, the diffusion coefficient in Zone A appears to be approximately one order of magnitude lower than the standard Galactic value ($\chi = 3.0$), which does not match the measured diffusion coefficient in Zone C. This result suggests the presence of anisotropic cosmic-ray diffusion in SNR G213 region.

\begin{figure}
     \centering
    \includegraphics[trim={0 0.cm 0 0}, clip,width=0.5\textwidth]{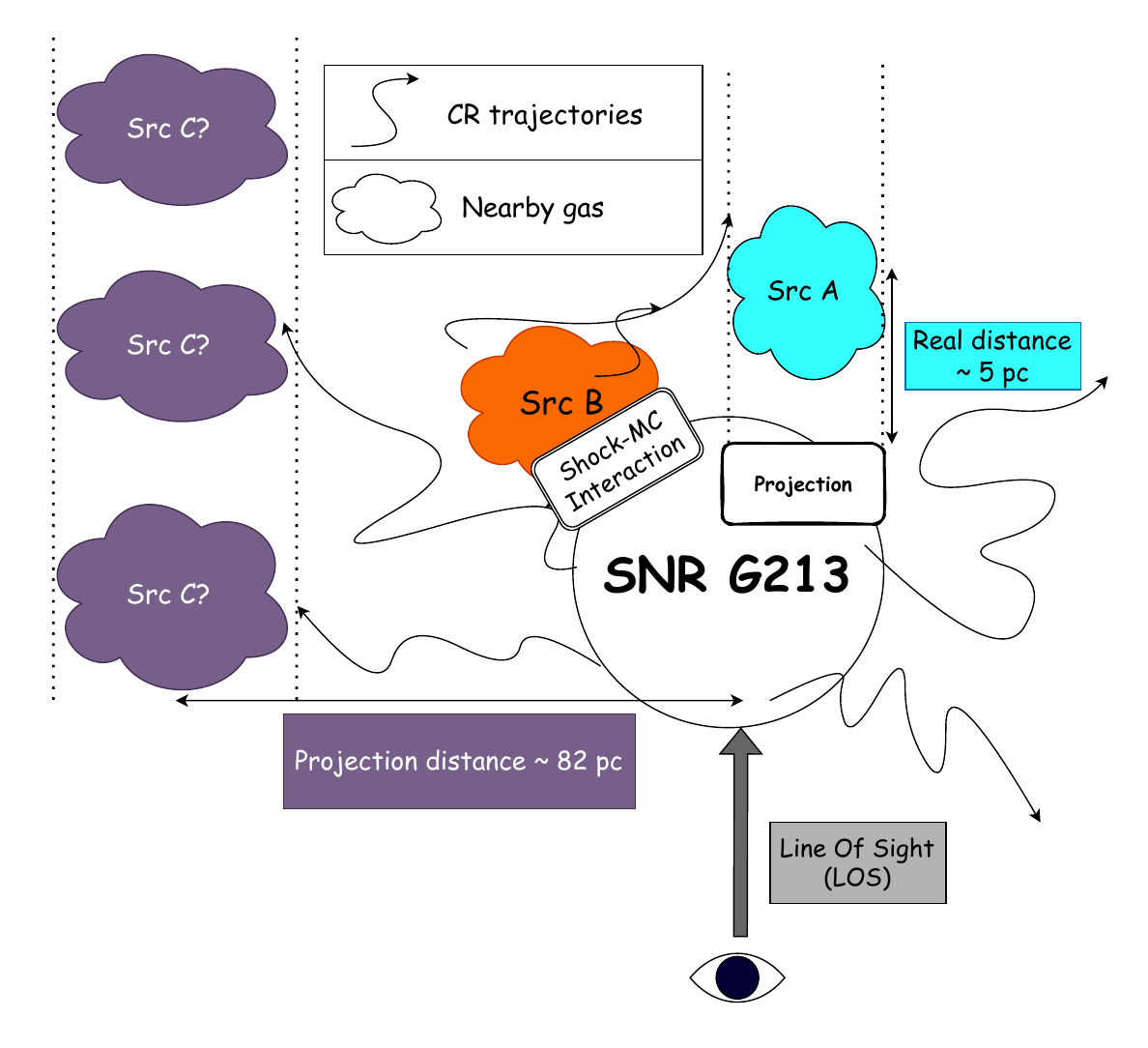}
    \caption{Sketch of the CR escaped from SNR G213 and illuminating nearby gas.}
\label{fig:5}
\end{figure}

The sketch of the above model can be seen in Figure \ref{fig:5}. In this scenario, if the SNR is associated with the gas in the velocity range of 35 - 54 km s$^{-1}$, then its real distance to Earth should be constrained around $D$ = 4.4 kpc. Since the projected angular distance between SrcC and SNR shell is $\theta$ = 1.07$\degr$, the corresponding physical distance is calculated as r$_{\rm s}$ = $\theta \times D$ = 82 pc. In contrast, for SrcA, which is located within the projected boundary of the SNR, the fit results in the left panel of Figure \ref{fig:4} suggest that it is located very close to the shock front.

\subsection{YSC scenario for SrcA at the distance around 1 kpc}\label{sec:4.2}
As mentioned above, young stellar clusters (YSCs) associated with star-forming regions (SFRs) can also contribute to the population of Galactic CRs \citep{2019NatAs...3..561A}. As shown in Figure \ref{fig:1}, SrcA is spatially coincident with a giant ionized region SH2-284 \citep{1976A&AS...25...25D}. In this region, several OB stars, multiple infrared sources and a maser source are all located inside SrcA. Also, the SFR G212.064-0.739 is reported to be embedded in several clusters by \citet{2012A&A...542A...3S}, for example Cl Dolidze 25, [FSR2007] 1069 and Cl Dolidze 49 \citep{2013A&A...558A..53K,2013MNRAS.436.1465B}, shown as green diamonds. Furthermore, more than 200 Young stellar object candidates and more than 100 Young stellar objects \citep{2003yCat.2246....0C,2020yCat.1350....0G} are detected, and the age of the inner stars was estimated to be more than 1 Myr \citep{2010A&A...509A.104D}, which is well consistent with previous $\gamma$-ray YSCs examples, such as Cygnus Coon \citep{2011Sci...334.1103A} and Westerlund 1 \citep{2012A&A...537A.114A}. Furthermore, the infrared survey confirmed 5 bubbles [HKS2019] E81,E82,E83,E84 and E85 in this region \citep{2019PASJ...71....6H}, indicating potential particle acceleration sites and the possibility of production of Galactic CRs\citep{1979ApJ...231...95M,1983SSRv...36..173C}. These evidence makes the assumption that the $\gamma$-ray emission might originate from the YSCs associated with the SFR reliable.

However, based on previous results, the OB stars are located in a wide range of distances, ranging from 0.6 to 5 kpc \citep{2021A&A...645L...8X,2024MNRAS.527.7355G}, and we found that the distance of this SFR also roughly matches with that of the maser source [PCC93] 111 \citep{1993A&AS..101..153P,1998A&AS..133..337Z}, of three open clusters: Cl Dolidze 25, [KPS2012] MWSC 0984, and [FSR2007] 1069 \citep{2013A&A...558A..53K,2013MNRAS.436.1465B}, and of several infrared sources detected by \citet{2009A&A...503..107P}, all of which are often thought as tracers of triggered star formation \citep{1998ASPC..148..150E,2005A&A...433..565D}. Considering all of them are located inside the R$_{\rm 68}$ boundary of SrcA, we assume here that part of OB stars are associated with the SFR and simply adopt the SFR distance to calculate the $\gamma$-ray flux of SrcA.

In this case, the total energy of protons for SrcA above 1 GeV is calculated as 3.5$\times$ 10$^{47}$ ($({n/52 \ cm^{-3})}^{-1} {(d/{1.0 \rm kpc}})^{2}$ erg, and the fitting results for a hadronic scenario is shown as the magenta line in the left panel of Fig \ref{fig:spectra}. This result is at least an order of magnitude lower than that of others YSCs \citep{2019NatAs...3..561A}. Based on the estimation method proposed by \citet{2004A&A...424..747P}, and considering that the number of detected OB stars within SrcA region is around 18 \citep{2024MNRAS.527.7355G}, the cumulated kinetic wind power from these massive stars can be calculated using typical wind parameters for OB-type stars. Assuming a standard mass-loss rate and typical wind velocity, the kinematic luminosity from a single massive star would range from 10$^{36}$ erg/s $\sim$ 10$^{37}$ erg/s \citep{1990A&A...232..119P,2001A&A...369..574V}. Considering most of the massive star within SrcA region are O-star with higher power, we adopted $L_{\rm wind} \sim$ 3$\times$ 10$^{36}$ erg/s \citep{2006MNRAS.367..763S} into following calculation, then the total kinetic wind power should be around 5.4$\times$ 10$^{37}$ erg/s, which is close to one-fourth of the kinetic luminosity of Cygnus Cocoon \citep{2011Sci...334.1103A}, while much lower than the Westerlund 1 with 1$\times$ 10$^{39}$ erg/s \citep{2006ApJ...650..203M}. Under the assumption of 1.0 kpc, the GeV $\gamma$-ray luminosity of SrcA in the range of [0.1,100] GeV is calculated to be 1.1 $\times$ 10$^{33}$ ${(d/{1.0 \rm kpc}})^{2}$ erg s$^{-1}$, which represents $\sim$ 0.002$\%$ of the stellar wind power. This value matches with Westerlund 1 \citep{2023A&A...671A...4H} and is slightly lower than that for Westerlund 2 with $\sim$ 0.005$\%$ \citep{2018A&A...611A..77Y} and near an order of magnitude  lower than Cygnus Cocoon with $\sim$ 0.03$\%$ \citep{2011Sci...334.1103A}. However, considering the measured R$_{\rm 68}$ of SrcA is 0.41$\degr$, under the 1 kpc distance, the physical extension size is around 8 pc, which is much smaller than others YSCs, for example, several tens pc for Westerlund 1 and Cygnus Cocoon \citep{2019NatAs...3..561A}. Furthermore, the typical lifetime of OB stars is of the order of a few million years, with an average value of T $\sim$ 5 Myr commonly adopted for star formation and feedback studies (e.g., \citep{1994ARA&A..32..227M,2000A&A...361..101M}). Under this assumption, the diffusion coefficient $\rm{D = r^2 / 4T}$ can be calculated as $\sim 9.6 \times 10^{23} \rm{cm^2 s^{-1}}$. For Bohm diffusion, its diffusion coefficient is $\rm{D_{B}} = c\times r_{\rm g} /3$ = 3 $\times 10^{24} \ \rm(E/ 1 TeV) \ \rm{(10\mu G/B) \ cm^2 s^{-1}}$, where r$_{\rm g}$ is the gyro radius. Thus for the 100 GeV protons responsible for 10 GeV $\gamma$-rays, the Bohm diffusion coefficient can be calculated as $\sim$ 3 $\times 10^{23} \rm {cm^2 s^{-1}}$ when an interstellar magnetic field B $\sim$ 10 $\mu$G is adopted. This estimated diffusion coefficient is only three times the Bohm limit, which may indicate the presence of CR-driven instabilities around the target source \citep{2013MNRAS.431..415B}. However, we note that the best-fit particle spectrum index for SrcA is $\sim$ 2.8 is much softer than the typical index of 2.2 $\pm$ 0.1 found in \citet{2019NatAs...3..561A}.

\section{Conclusions}\label{sec:5}

We analyzed the GeV $\gamma$-ray emission in the vicinity of SNR G213.0-0.6 using 16 yr of Fermi-LAT data. Three extended $\gamma$-ray sources were detected, named as SrcA, SrcB and SrcC, respectively. SrcC is located outside the southeastern edge of the SNR G213, while SrcA and SrcB are located inside the radio shell. Three dense molecular clouds (MC) are detected in this region based on the MWISP CO observation. SrcB is spatially coincident with one of the MC and the SNR radio shell, which supports the hypothesis that SrcB is the GeV counterpart of SNR G213.0-0.6. The spatial coincidence between MC and SrcA, SrcC suggests that their $\gamma$-ray emission could originate from the hadronic $\pi^0$ decay due to the collisions between the escaped CRs and the protons inside the MC. In our model, the injected proton spectra for each source are kept the same, but the $\gamma$-ray spectrum from SrcA and SrcC could be explained with different parameters in the escape model, e.g. the diffusion coefficient or the escape distance of the CRs. For SrcA, the escape distance r$_{\rm s}$ is relatively small, and the $\chi$ value is lower than the Galactic standard value, indicating a potentially slow diffusion. For SrcC, the $\gamma$-ray spectrum can be well fitted by a larger diffusion coefficient close to the Galactic standard value, with a longer escape distance. For SrcB, the $\gamma$-ray spectrum is consistent with the hadronic $\pi^0$ decay scenario for protons with a single power-law spectrum with an index $\sim$ 2.5.

Furthermore, we also discuss the possibility that the $\gamma$-ray emission of SrcA originates from the CRs accelerated by the YSCs associated with a nearby SFR G212.064-0.739, which includes several OB stars, infrared sources and one maser source, all of them being located within R$_{\rm {68}}$ of SrcA, and inside the ionized hydrogen region SH 2-284. In this scenario, the calculated proton energy above 1 GeV is much lower than other typical $\gamma$-ray-emitting YSCs examples (e.g. Westerlun 1, Westerlund 2), while the calculated conversion efficiency of stellar wind is close to these known $\gamma$-ray-emitting YSCs. Considering the 1 kpc distance of the SFR, and adopting an average age for these OB stars, the derived diffusion coefficient ($\rm{D = r^2 / 4T}$) is approximately three times the Bohm limit. This may indicate the presence of CR-driven instabilities generating a suppressed diffusion region around the target source. However, the current evidence is still insufficient to draw a solid conclusion. Further observations with high sensitivity detectors like CTA \citep{2013APh....43....3A} and LHAASO \citep{2019arXiv190502773C} towards this region are warranted.


\begin{acknowledgements}
This research made use of the data from the Milky Way Imaging Scroll Painting (MWISP) project, which is a multi-line survey in 12CO/13CO/C18O along the northern galactic plane with PMO-13.7m telescope. We are grateful to all the members of the MWISP working group, particularly the staff members at PMO-13.7m telescope, for their long-term support. MWISP was sponsored by National Key R$\&$D Program of China with grants 2023YFA1608000 $\&$ 2017YFA0402701 and by CAS Key Research Program of Frontier Sciences with grant QYZDJ-SSW-SLH047. We also would like to thank Xuyang Gao, Xiaona Sun, Yang Su and P.P.Delia for invaluable discussions. This work is supported by the National Natural Science Foundation of China under the grants No. 12393853, U1931204, 12103040, 12147208, and 12350610239, the Natural Science Foundation for Young Scholars of Sichuan Province, China (No. 2022NSFSC1808), and the Fundamental Research Funds for the Central Universities (No. 2682022ZTPY013).
\end{acknowledgements}

\bibliography{ref}
\bibliographystyle{aasjournal.bst}

\end{document}